\documentclass[amsmath,amssymb, %preprint,% 
reprint,%
]{revtex4-1}

\usepackage[pdftex]{graphicx}% Include figure files
\usepackage{color}
\usepackage{transparent}
\usepackage{dcolumn}% Align table columns on decimal point
\usepackage{bm}% bold math
\usepackage{epstopdf}

\begin{document}

\title[ColeArxiv]{Label-free single molecule imaging with numerical aperture-shaped interferometric scattering microscopy}

\author{D. Cole$^*$}
\affiliation{Physical and Theoretical Chemistry Laboratory, University of Oxford, South Parks Road, OX1 3QZ Oxford, UK.}
\author{G. Young}
\email{These authors contributed equally to this work.}
\affiliation{Physical and Theoretical Chemistry Laboratory, University of Oxford, South Parks Road, OX1 3QZ Oxford, UK.}
\author{A. Weigel}
\altaffiliation{Current address: Ultrafast Innovations, Am Coulombwall 1, 85478 Garching, Germany.}
\affiliation{Physical and Theoretical Chemistry Laboratory, University of Oxford, South Parks Road, OX1 3QZ Oxford, UK.}
\author{P. Kukura}
\email{philipp.kukura@chem.ox.ac.uk}.
\affiliation{Physical and Theoretical Chemistry Laboratory, University of Oxford, South Parks Road, OX1 3QZ Oxford, UK.}


\maketitle

\textbf{Our ability to optically interrogate nanoscopic objects is controlled by the difference between their extinction cross sections and the diffraction-limited area to which light can be confined in the far-field. Here, we show that a partially transmissive spatial mask placed near the back focal plane of a high numerical aperture microscope objective enhances the extinction contrast of a scatterer near an interface by approximately $T^{-1/2}$, where $T$ is the transmissivity of the mask. Numerical aperture-based differentiation of background from scattered light represents a general approach to increasing extinction contrast and enables routine label-free imaging down to the single molecule level.}

%Add 5-10 citations in first paragraph

Ultrasensitive optical microscopy has transformed our ability to investigate structure and dynamics on microscopic and nanoscopic length scales.\cite{Kulzer:2004by, Walt:2012ko, Weisenburger:2015jb} Among the many available contrast mechanisms, fluorescence microscopy has been the method of choice for the detection and imaging of single molecules, due to the extremely high achievable imaging contrast.\cite{Orrit1990,Weiss:1999tf,Moerner:2003ho} Imaging non-fluorescent species at comparable sensitivity levels, however, remains challenging due to the difficulty associated with differentiating light that has interacted with the object of interest from any background.\cite{Arroyo:2015kr} Recently, resonant detection of single molecules has been reported using a number of approaches,\cite{Kukura:2010cy,Chong:2010bc,Gaiduk:2010bf} while non-resonant detection and imaging has been demonstrated by interferometric scattering microscopy (iSCAT).\cite{Arroyo:2014bh,Piliarik:2014dp} 

Extinction detection in its most traditional implementation relies on the  difference in light transmitted by a sample in the presence and absence of the object of interest. It has been shown that in extreme cases a single molecule \cite{Kukura:2010cy} or even an atom can be detected in this fashion.\cite{Utikal:2014bd} For a single molecule under ambient conditions, the interference contrast when measuring in transmission amounts to a few parts-per-million, calling for sophisticated noise suppression.\cite{Kukura:2010cy} The interferometric contrast increases when the experiment is performed in reflection at the expense of a reduction in the detected photon flux, leading to an identical signal-to-noise ratio when using the same incident light intensity, extinction cross section and integration time for shot noise-limited detection.\cite{OrtegaArroyo:2012kk,Arroyo:2016iz} For an extinction experiment performed in a reflective arrangement as in iSCAT, the detected light intensity can be written as:
\begin{equation}
I_{\textrm{det}}=|E_{\textrm{inc}}|^2\left[r^2+|s|^2+2r|s|\textrm{cos}\phi\right]
\end{equation}
where $E_{\textrm{inc}}$ denotes the incident electric field, $r^2$ the reflectivity of the sample or interface, $s$ the scattering amplitude of the object of interest, and $\phi$ the phase difference between scattered and transmitted fields at the detector. For weak scatterers, $|s|^2$ becomes negligible compared to the other contributions, reducing the iSCAT contrast,  defined as the ratio in the detected light intensities in the presence, $I_{\textrm{det}}$, and absence, $I_{\textrm{bkg}}$, of a scatterer to:
\begin{equation}
\frac{I_{\textrm{det}}}{I_{\textrm{bkg}}}=1+\frac{2|s|\textrm{cos}\phi}{r}.
\end{equation}

For extinction detection in biologically compatible environments, the ratio between the reflectivity, usually determined by the refractive indices of standard microscope coverglass (n=1.52) and water (n=1.33), combined with the extinction cross sections of single proteins, results in iSCAT contrasts on the order of $0.4\%/$MDa of molecular mass for illumination in the near ultraviolet.\cite{Arroyo:2014bh,Piliarik:2014dp} Detecting signals that amount to $0.1\%$ or less of the background light requires detection of $10^8$ photoelectrons for a signal-to-noise ratio of 10 even for a perfectly executed experiment where shot noise-induced fluctuations of the background are the only noise source. From an imaging perspective, this is experimentally challenging, since almost all commercially available digital cameras do not provide full well capacities beyond $10^5$ photoelectrons, requiring significant temporal and spatial averaging to detect the required number of photons.\cite{Kukura:2009eb} Furthermore, the strong focus dependence of the iSCAT signal caused by Gouy phase contributions to $\phi$\cite{Lindfors:2004bo} makes it difficult to determine the correct focus position and thus visualize weak scatterers, such as single proteins, since they are usually only revealed in post-processing of the acquired data. 

An attractive approach to enhance the contrast of weak scatterers would thus involve reducing the reflectivity of the interface as can be immediately seen from Eq. 2. In principle, this could be achieved by appropriately coating the substrate surface, although the scope here is limited due to residual reflections from lenses within microscope objectives and the cost associated with specifically coated substrates. An alternative route to improving the ratio between scattered and background, or reflected light involves taking advantage of the different directionalities of illumination and scattered light in a combination of dark field and interference reflection microscopy (Fig. 1).\cite{Curtis1964} As has been shown previously,\cite{Lieb:2004kr,Brokmann:2005de, Lee:2011cr} a point source near a refractive index interface radiates the majority of photons into the higher index material in directions associated with a high numerical aperture of the collecting lens (Fig. 1a). At the same time, widefield illumination in optical microscopes is usually achieved by focusing an incident beam into the back focal plane of the imaging objective, which requires the use of only a very low numerical aperture. As a result, it is possible to very effectively discriminate between scattered and illumination light, suppressing the illumination light by more than 7 orders of magnitude with little loss in the detection efficiency of scattered light.\cite{Weigel:2014fw} 

In its simplest implementation, such an approach uses a small reflective mirror near the back aperture of the imaging objective to efficiently couple the illumination light in and out of the microscope (Fig. 1b), which for a fully reflective mirror leads to efficient dark field imaging.\cite{Sowa:2010cf} If the mirror, however, consists of a thin metallic layer coated onto a glass window, a fraction of the illumination light reflected by the glass-water interface will pass through the reflective layer and reach the camera together with any scattered light collected by the objective, which passes the mirror largely unaffected. 

The effect of the partial reflector is to reduce the background light intensity reaching the detector, thereby increasing the interferometric contrast. The degree of attenuation can be tailored to the incident light intensity and camera properties by simply changing the thickness of the metallic layer. The reduction in scattered light detection caused by this experimental arrangement, however, only amounts to $11.3\%$ when using an oil immersion objective with an 8.52 mm diameter back aperture in combination with a 3.5 mm diameter partially reflective mirror (Fig. 1c).\cite{Weigel:2014fw} As a result, the partial reflector selective attenuates reflected over scattered light.

\begin{figure}[htbp]
\includegraphics{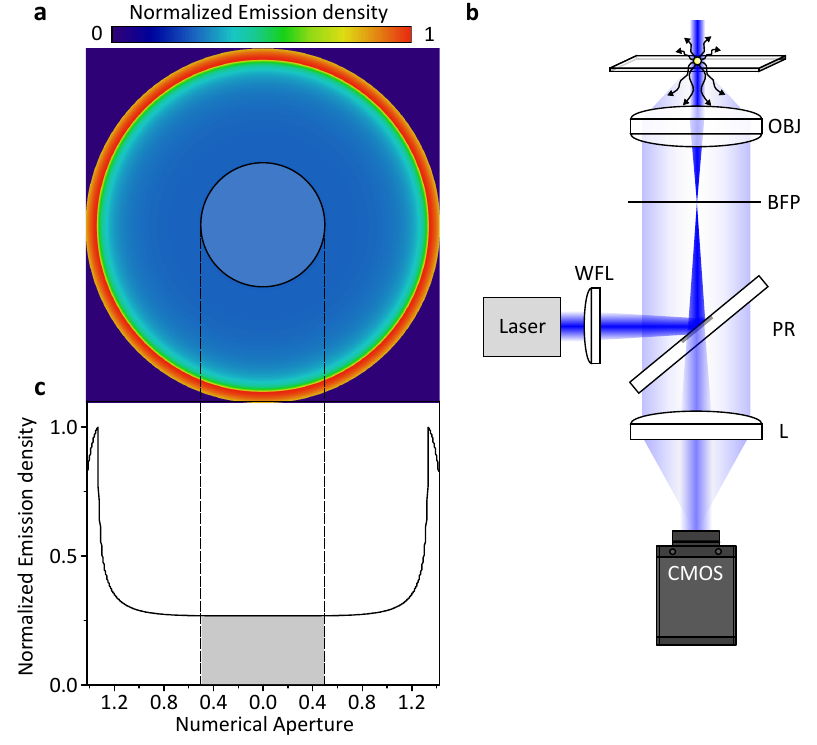}
\caption{\label{fig:fig1} Concept and experimental realization of numerical aperture-filtered interferometric scattering microscopy (iSCAT. (a) Emission pattern of a nanoscopic scatterer at a glass-water interface emerging from the back aperture of a high numerical aperture (1.42) microscope objective for circularly polarized illumination.\cite{Lieb:2004kr} The semitransparent circle indicates the region occupied by a partial reflector shown in the schematic experimental setup in (b). Here, a single mode fiber-coupled diode laser is loosely focused with a wide field lens (WFL) into the back focal plane (BFP) of an oil immersion objective (OBJ). The partial reflector (PR) couples illumination light in and out of the inverted microscope and attenuates the reflected light before being imaged onto the camera (CMOS) by another lens (L). (c) Emission density as a function of numerical aperture, with the gray area indicating the region attenuated by the partial reflector.}
\label{fig1}
\end{figure}

To illustrate the effect of numerical aperture filtering on the interference contrast, we begin by comparing images of bare microscope cover glass immersed in water acquired with a simple iSCAT microscope\cite{Piliarik:2014dp} (Fig. 2a,b) with the setup shown in Fig. 1b using a partial reflector (Fig. 2c,d) which transmits approximately $1\%$ of the returning reflection from the glass coverslip. For standard iSCAT imaging, the image of a loosely focused beam (5 $\mu$m full width at half maximum) reflecting off the surface of the cover glass appears largely featureless. Scattering contrast caused by cover glass roughness on the order of $2-4\%$ of the total detected light intensity only becomes visible after division by a median image removing illumination inhomogeneities.\cite{Arroyo:2016iz} Use of the spatial mask on the other hand enhances the scattering contrast by approximately a factor of 10 (Fig. 2d), making it already visible in the raw camera image (Fig. 2c).

\begin{figure}[htbp]
\includegraphics{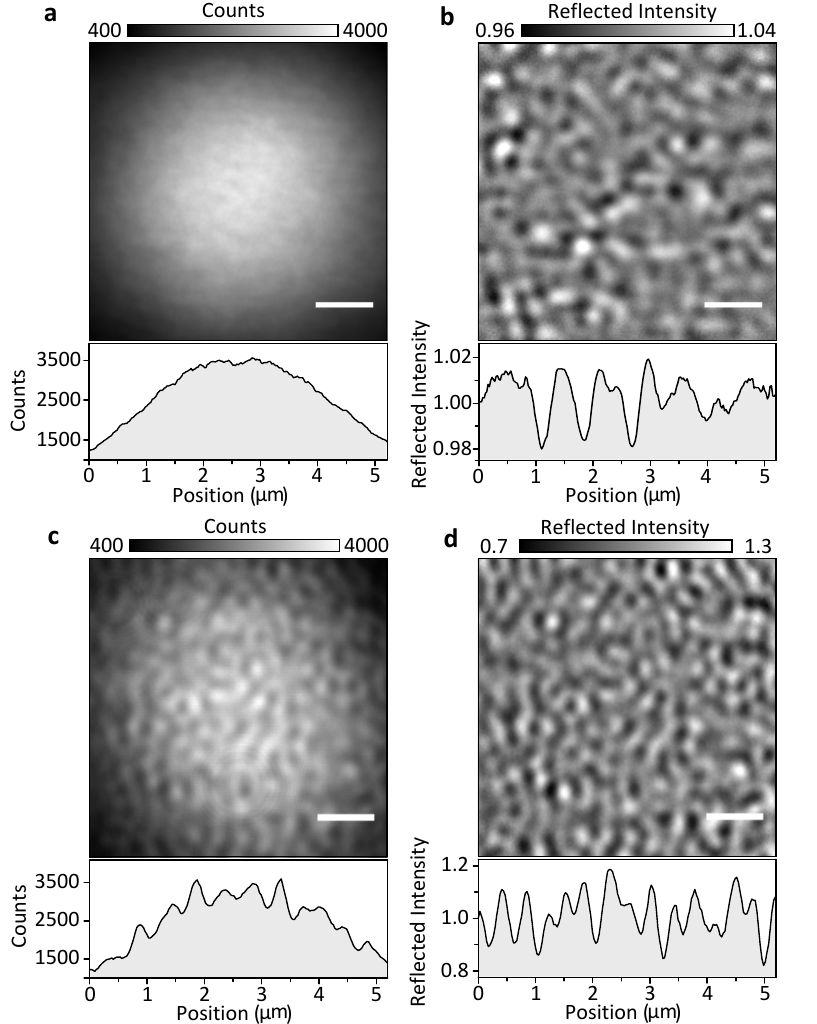}
\caption{\label{fig:fig2} Effect of a partial reflector on iSCAT contrast. (a) Raw reflection image of microscope cover glass covered by water upon illumination with a 445 nm laser using a combination of polarising beam splitter and quarter wave plate to separate incident from reflected and scattered light. (b) The same image as in (a) after division by a median image produced while laterally translating the sample.\cite{Arroyo:2016iz} (c,d) Equivalent images taken with the setup shown in Figure 1 detailed in the Methods. To generate comparable images, we used a 100-fold longer exposure time for (c) compared to (a), while (a) represents the average of 100 consecutive images. In this way, the total number of incident and detected photons is identical for (a) and (c). Scale bars: 1 $\mu$m.}
\label{fig2}
\end{figure}

We remark that use of the mask with shot noise limited sensitivity does not improve the signal-to-noise ratio of interferometric scattering detection, as it increases the scattering contrast at the expense of the detected photon flux. Given that the mask reduces the incident light intensity on the camera 100-fold, the shot noise-induced background fluctuations for the same exposure time and incident light intensity increase by $\sqrt{100}=10$, which matches the ratio between the observed scattering contrasts. For shot noise-limited detection of the same incident photon flux over the same integration time, any illumination and detection approach, including transmission, will thus yield similar signal-to-noise ratios. 

The increased scattering contrast enabled by the partial reflector at the expense of the total detected photon flux, however, has two important consequences. Firstly, it enables the use of both high speed and high quantum efficiency cameras with low full well capacities, enabling optimal use of the incident and scattered photons, without having to resort to large magnifications or temporal averaging.\cite{Kukura:2009eb} As a result, surface roughness of microscope cover glass now becomes visible in single images acquired using standard digital cameras with full well capacities on the order of few $10^4$ photoelectrons leading to shot noise induced fluctuations on the order of $<1\%$ root mean squared (RMS). An additional advantage is that the enhanced contrast simplifies the otherwise challenging task of determining optimal focussing conditions for label-free iSCAT, which is best performed in the absence of any strong scatterers on an ideally defect-free surface.   

The remaining challenge associated with detecting very weak scatterers such as unlabeled single proteins with iSCAT is that the respective signals are significantly smaller than those produced by the rough cover glass substrate, requiring efficient background removal.\cite{Arroyo:2014bh,Piliarik:2014dp} In a simple landing assay, where single molecules bind to and unbind from a surface, individual events can be revealed by  subtracting images prior to a binding event from those following it. An intrinsic problem that arises, however, is that the precise binding time cannot be controlled and may thus occur during a single camera exposure. If individual, subsequent frames are subtracted, this can lead to significant blurring of the perceived iSCAT contrast, limiting the precision of the extinction contrast measurement. 

To avoid such inaccuracies, we found it optimal to design the transmissivity of the mask such as to reach near saturation of the imaging camera when using the highest desirable incident power at the fastest possible read out speed for our field of view ($5\times5$ $\mu$m). In principle, a lower transmissivity can be used to increase the interferometric contrast further, although care has to be taken to avoid increasing relative contributions from background light, such as residual reflections from the objective, which can reduce the interferometric contrast. 

The apparent temporal oversampling inherent to our approach, enables us to precisely determine  individual (un)binding events in time. To achieve this, we record an iSCAT movie (Supplementary Movie 1) at the highest frame rate enabled by our camera before dividing normalized averages of several tens of frames to produce a ratiometric image. We then repeat this process shifting one frame forward at a time, producing a new movie now displaying the resulting ratiometric images (Fig. 3a and Supplementary Movie 2). The timing of the scatterer binding to the surface relative to the midpoint between the two frame batches producing the ratiometric image then determines the scattering contrast. 

For a simple single molecule binding assay, the resulting ratiometric images invariably exhibit a range of scattering contrasts (Fig. 3b) due to the random binding events in time. This is true even for a highly uniform sample, in our case the small heat-shock protein from \textit{Methanocalcocuccus janaschii} HSP16.5, which forms a 24-mer with total molecular mass of 396 kDa.\cite{Kim1998a} The accurate scattering contrast for each event is revealed when the binding event coincides precisely with the mid-point marking the division between the two frame batches (Fig. 3c). The resulting movie thus appears to show objects slowly appearing and disappearing (Supplementary Movie 2), but the associated contrast histogram exhibits a Gaussian distribution as expected for a shot noise limited detection process (Fig. 3d).
Given that such an image analysis approach can be easily implemented for video rate imaging, single molecule binding events become visible in real time upon using the data processing approach described above, further simplifying label-free single molecule detection. Importantly, the use of a running image ratio removes the necessity for extreme control of sample drift, which is required for a static approach.

\begin{figure}[htbp]
\includegraphics{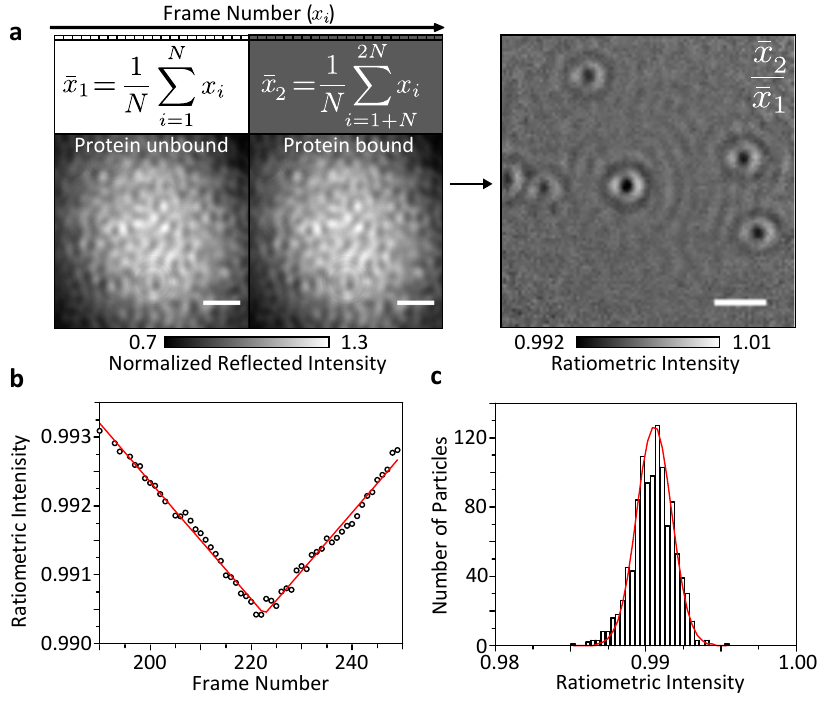}
\caption{\label{fig:fig3} Ratiometric iSCAT imaging for detection of weak scatterers. (a) Concept of iSCAT ratiometric imaging dividing averages of consecutively recorded images. (b) Resulting ratiometric image using N=100 revealing the binding of HSP16.5 to a bare cover glass surface. Incident power density: 45 kW/cm$^2$; exposure time: 1.4 ms; frame rate: 700 Hz; $2\times 2$ pixel binning; HSP16.5 monomer concentration: 240 nM; scale bar: 1 $\mu$m. (c) iSCAT contrast as a function of frame number dividing the two sets of averaged frames used for ratiometric imaging in (a). (d) Histogram of signal contrasts for 1083 events.}
\end{figure}

The simplicity afforded by the partial reflector comes at the price of some disadvantages compared to previous implementations of iSCAT, in particular those based on rapid beam scanning.\cite{Kukura:2009eb,Arroyo:2014bh} We found that the imaging background caused by cover glass roughness is relatively enhanced compared to scatterers bound to the surface, likely due to the exclusion of low frequency components by the partial reflector, which can be seen in the comparison of Figs. 2b and 2c. Using effectively wide-field illumination also leads to extended Airy rings visible in Fig. 3b, which cause significant image distortions for larger scatterers, such as gold nanoparticles or microtubules (Supplementary Figure 1).  Furthermore, focusing the illumination light into the back focal plane of the objective results in an effectively collimated beam traveling through the sample, thus maintaining a high power density along the optical axis. This increases imaging background from light scattered by species diffusing in solution when compared with an approach where a beam is comparatively tightly focused.\cite{Kukura:2009eb} Finally, the ultimate sensitivity for the setup shown in Fig. 1b was limited by the appearance of background features we believe to be originating from laser noise, either caused by frequency or beam pointing drifts (Supplementary Movie 2). 

Many of these shortcomings can be addressed by combining the concepts presented here with the original illumination approach used by iSCAT (Fig. 4a).\cite{Kukura:2009eb, Arroyo:2016iz} Placing the partial reflector in an image plane in the detection channel enables direct access to the back focal plane of the objective and thereby optimizes the attenuation of the scanned illumination beam. This approach produces iSCAT images of unlabeled proteins with unprecedented clarity, albeit at the expense of increased experimental complexity compared to the setup shown in Fig. 1b. For a 396 kDa protein recorded at 3 Hz, we find a signal-to-noise ratio of 23 defined as $R/\sigma$, where $R$ is the ratiometric intensity and $\sigma$ the RMS fluctuations of the background. In particular, scanning a focused illumination beam reduces extended diffraction rings and imaging background caused by diffusing scatterers and laser noise (Fig. 4b). Using this approach, we found the label-free imaging sensitivity of iSCAT to be exclusively limited by sample drift, which causes features from glass roughness to bleed into the ratiometric images (Fig. 4c). The interference contrast for HSP16.5 slightly decreases compared to widefield illumination (Fig. 3c) due to an increase in background light caused by the use of a collimated rather than a focused beam in the back focal plane of the objective.

\begin{figure}[htbp]
\includegraphics{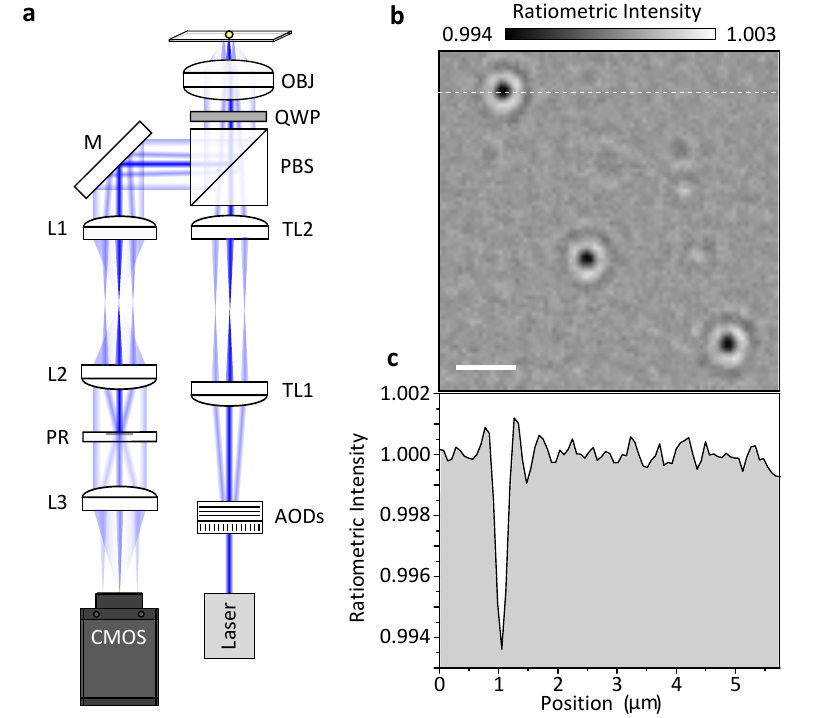}
\caption{\label{fig:fig4} Numerical-aperture filtered iSCAT using rapid beam scanning (a) Experimental setup based on acousto optic (AOD) scanning of a collimated beam using a telecentric imaging setup (TL1/2). The combination of a $\lambda/4$ waveplate (QWP) and polarising beam splitter (PBS) separates incident from scattered and reflected light. Two lenses (L1/2) create an image of the back focal plane of the microscope objective (OBJ) where the partial reflector (PR) attenuates the backreflected beam while leaving scattered light effectively unchanged. A final lens (L3) forms an image on the CMOS camera. Using a 3.5 mm diameter partial reflector together with a 1 mm diameter collimated illumination beam provides a good compromise between confocal sectioning and background attenuation. (b) Resulting ratiometric image using N=150 revealing the binding of HSP16.5 to a bare cover glass surface. The weak bright features are caused by proteins unbinding from the surface. Incident power density: 150 kW/cm$^2$; exposure time: 1.9 ms; frame rate: 500 Hz; $3\times 3$ pixel binning HSP16.5 monomer concentration: 240 nM; scale bar: 1 $\mu$m. (c) Image cross section from the dashed white line shown in (b).}
\end{figure}

The presented approach allows for facile label-free single molecule detection with almost any inverted light microscope equipped with a digital camera and a high numerical aperture oil immersion lens. Coupling illumination light in and out of the imaging path can be easily miniaturized\cite{Ishmukhametov:2016hf} paving the way for widespread adoption not only on custom-built but also on commercial microscopes. The reduction of light intensity reaching the imaging camera enables use of essentially any digital camera, without requirement for high full well capacity or low read noise. We also found that the use of the spatial mask allowed for large fields of view up to $20\times20$ $\mu$m without requiring any beam scanning (Supplementary Figure 2). 

In addition to these simplifications, we have shown that the partial reflector can be used to increase the interferometric contrast in a more standard iSCAT experimental arrangement. This primarily enables the use of higher incident powers without suffering from camera saturation or having to resort to inefficient, expensive and complex cameras or large magnifications that are incompatible with commercial microscopes. Since the ultimate limits for iSCAT detection and imaging are defined by shot noise-induced background fluctuations, higher incident powers automatically translate into higher detection sensitivity and precision. The possibility of some increases in incident power density, given that phototoxicity and heating are of little consequence for single molecule binding assays, together with increased temporal averaging as enabled by more stable experimental setups\cite{Walder:2015fv} should lead to improved detection sensitivities and precision down to a few kDa $\times$ Hz$^{1/2}$ without the need for the development of completely new camera technologies, thereby paving the way towards single molecule mass spectrometry in solution and robust label-free single molecule detection and imaging using iSCAT.
\section*{Acknowledgments}
P.K. was supported by an ERC Starting Investigator Grant (Nanoscope, 33757) and the EPSRC (EP/M025241/1). G.Y. was supported by a Zvi and Ofra Meitar Magdalen Graduate Scholarship. We thank Olga Tkachenko and Justin Benesch for providing HSP16.5. 

\section*{Methods}
The experimental setup used for collecting the data shown in Fig. 2 is similar to that reported previously\cite{Piliarik:2014dp} apart from the fact that illumination light is separated from scattered and reflected light by our partially reflective mirror rather than a beamsplitter. Briefly, we loosely focus the output of a fiber-coupled 445 nm diode laser (Lasertack) into the back focal plane of an oil immersion objective (Olympus, 1.42 NA, 60x) and couple it in and out of the imaging path using our partial reflector that is evaporated onto a thin glass window placed 2 cm from the entrance pupil of the objective. A 600 mm focal length lens images scattered and reflected light onto a CMOS camera (Point Grey GS3-U3-23S6M-C) at 200x magnification. The focus position was stabilized with an active feedback loop. The partially reflective mirror consists of a 82 nm thick silver layer of 3.5 mm diameter evaporated onto a 3 mm thick glass window. We emphasize that in the experiments reported here, we used a circular mask, which results in an elliptical attenuation when placed at 45 degrees (Fig. 1b), but found the influence of an elliptical over a spherical mask in terms of the image quality negligible as evidenced by the comparison of Figs. 3a and 4b. The experimental setup used for Fig. 4 is identical to the one described in detail\cite{Arroyo:2016iz} except for the introduction of the partial reflector, an intermediate image plane and use of the CMOS camera identified above.

\bibliography{Cole_Refs}% Produces the bibliography via BibTeX.

\end{document}